\documentclass[aps,prl,reprint,superscriptaddress]{revtex4-2}

\usepackage{graphicx}
\usepackage{dcolumn}
\usepackage{bm}

\usepackage[english]{babel}

\usepackage{amsmath}

\usepackage{physics}
\usepackage{braket}
\usepackage{siunitx}

\usepackage{standalone}

\usepackage{hyperref}
\hypersetup{colorlinks,allcolors=blue}
\usepackage{tikz}
\usetikzlibrary{calc}
\usetikzlibrary{positioning}

\usepackage{pgfplots}
\usepgfplotslibrary{colorbrewer}
\usepgfplotslibrary{colormaps}

\pgfmathsetlengthmacro\MajorTickLength{
  \pgfkeysvalueof{/pgfplots/major tick length} * 0.7}

\begin{document}

\title{Programmable coherent site-selective spin control in rotating Penning-trap ion crystals}

\author{Nihar Makadia}

\email[e-mail: ]{nihar.makadia@sydney.edu.au}
\author{Julian Y. Z. Jee}
\author{Gustavo Caf\'e de Miranda}
\author{Joseph H. Pham}
\author{Michael J. Biercuk}
\affiliation{School of Physics, The University of Sydney, Sydney, NSW 2006, Australia}

\author{Athreya Shankar}
\affiliation{Department of Physics, Indian Institute of Technology Madras, Chennai, 600036, Tamil Nadu, India}

\affiliation{Center for Quantum Information, Communication and Computing, Indian Institute of Technology Madras, Chennai, 600036, Tamil Nadu, India}

\author{Robert N. Wolf}
\affiliation{School of Physics, The University of Sydney, Sydney, NSW 2006, Australia}
\affiliation{Sydney Nano Institute, The University of Sydney, Sydney, NSW 2006, Australia}

\date{\today}

\begin{abstract}
    Large ion crystals in Penning traps provide a platform for quantum simulation and sensing with hundreds of spins, but their continuous rigid-body rotation has so far limited flexible local qubit control. Here we demonstrate programmable coherent site-selective spin control across large rotating ${}^{9}\mathrm{Be}^{+}$ crystals in a Penning trap. A tightly focused off-resonant laser beam drives local $R_z$ phase rotations via differential AC Stark shifts. Beam steering synchronised with crystal rotation enables addressing of arbitrary ions throughout the crystal. Ramsey-based characterisation shows $R_z(\pi)$ gate fidelity of $0.950(4)$ and nearest-neighbour crosstalk error of $0.021(6)$. We demonstrate preparing spatially structured spin patterns, generating a biskyrmion spin texture in a single-layer crystal, and then extend the method to bilayer crystals where we perform layer-selective addressing operations. We further demonstrate dual-quadrature Ramsey sensing by imprinting a relative $\pi/2$ phase shift between spatial sub-ensembles, enabling simultaneous measurement of orthogonal spin components within a single experimental realisation. These results establish programmable local control in large rotating ion crystals, opening new routes for engineering spatially structured quantum states in multidimensional trapped-ion systems.
\end{abstract}
\maketitle

\section{Introduction}

Trapped-ion crystals are a powerful platform for quantum simulation, sensing, and computation due to their long coherence times, programmable interactions, and high-fidelity state preparation and measurement \cite{Bruzewicz2019,Monroe2021,Gilmore2021a,Lewis-Swan2019,Valahu2025}.
One-dimensional (1D) ion strings in Paul traps have enabled precise single-ion control and high-fidelity entangling gates, leading to major advances in quantum computation and simulation \cite{Debnath2016,Kokail2019,Nam2020,FossFeig2025}.
Complementing this approach, Penning traps naturally support rotating two-dimensional (2D) Coulomb crystals containing tens to hundreds of ions, enabling quantum simulation and sensing with large ion arrays \cite{Mitchell1998,Britton2012,Safavi-Naini2018,Bullock2026,Jee2026}. Recent proposals further extend this architecture to bilayer and multilayer ion crystals, opening new opportunities for quantum information processing in three dimensions \cite{Hawaldar2024}.
More recently, 2D ion crystals have also been realised in Paul traps, significantly increasing their ion number \cite{DOnofrio2021,Kiesenhofer2023,Guo2024,Qiao2024,Sun2024a}.

A central outstanding challenge, however, has been the realisation of coherent single-ion control across large 2D crystals, which is required for programmable preparation and manipulation of spatially structured quantum states \cite{DiVincenzo2000,Shankar2022a,Kaubruegger2025}.
While coherent single-ion control is well established in 1D ion strings and has recently been demonstrated in small 2D Paul-trap crystals \cite{Hou2024a}, extending this capability to large Penning-trap crystals has remained an outstanding challenge due to the intrinsic rotation of the crystal, which keeps the ions in continuous motion in the laboratory frame.
Two complementary approaches are currently being developed to address this challenge. One employs globally illuminating laser beams to generate spatially varying AC Stark shift patterns using adaptive optical wavefront shaping \cite{Polloreno2022,Carter2026} or engineered optical phase profiles \cite{Jee2026}. The other uses tightly focused laser beams to address individual ions directly, first demonstrated through a stationary focused AC Stark shift beam \cite{McMahon2024a}, and later for radius-selective optical repumping to prepare domain-wall-like spin structures \cite{Jee2026}.
\vspace{4pt}

Here, we demonstrate programmable and coherent control of single ions across a rotating  ${}^{9}\mathrm{Be}^{+}$ crystal in a Penning trap. Using a directable, tightly focused off-resonant laser beam, we implement site-selective $R_z$ phase rotations through differential AC Stark shifts on the qubit transition.
By combining single-shot ion localisation with synchronised beam steering and power modulation, we coherently address about 250 ions within a few hundred microseconds.
We demonstrate the versatility of this capability through layer-selective addressing in bilayer crystals \cite{Hawaldar2024}, deterministic preparation of a biskyrmion spin texture \cite{Shankar2022a, Kaneda2016, Shengbin2023, Psaroudaki2021, Koshibae2015, Luo2021}, and spatially resolved dual-quadrature Ramsey interferometry, in which orthogonal phase are measured simultaneously using spatially separated ion crystal sub-ensembles \cite{Li2022b,Zheng2024,Kaubruegger2025}. These results establish coherent local spin control as a key capability for large-scale Penning-trap quantum simulators and sensors.

\vfill

\section{Experimental system and procedure}

Laser-cooled ${}^{9}\mathrm{Be}^{+}$ ions are confined in a high-optical-access Penning trap \cite{Ball2019, Pham2024}, forming a rotating Coulomb crystal with an axial centre-of-mass mode frequency of $\omega_\mathrm{COM}/2\pi=\SI{698}{\kilo\hertz}$.
The crystal rotation frequency, $\omega_r/2\pi$, which is \SIrange{79}{85}{\kilo\hertz} in this work, as well as the crystal geometry, are controlled by a rotating-wall potential \cite{Huang1998a}. This leads to an average ion-ion spacing of $s\approx\SI{22}{\micro\metre}$ in 2D crystals. Site-resolved spin state readout in this rotating system is enabled by a spatially resolving single-photon timestamping camera \cite{Nomerotski2017}. Reconstructing photon positions in the frame co-rotating with the ions yields single-shot ion localisation and enables spin-state discrimination across large crystals \cite{Wolf2024}, which is essential both for reconstructing spatial spin textures \cite{Jee2026} and for calibrating the single-ion addressing sequence.

\begin{figure}[t!t]
    \centering
    \includegraphics{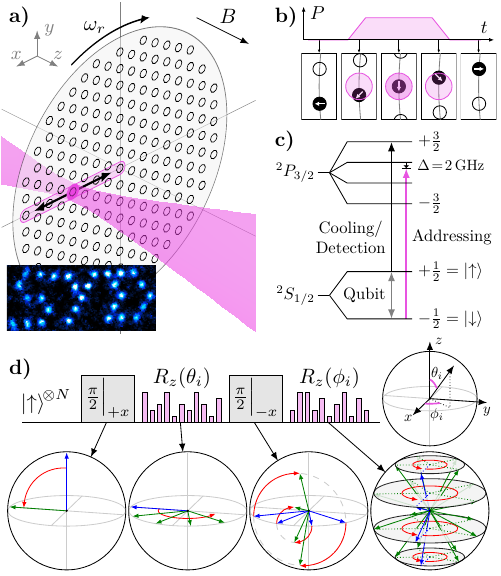}
    \caption{
        Experimental setup and protocol for programmable coherent site-selective spin control in a rotating Penning-trap crystal.
        (a) A focused off-resonant laser beam drives site-selective $R_z$ spin rotations by introducing a differential AC Stark shift ($\delta_\text{ACSS}$) on the qubit levels of ${}^{9}\mathrm{Be}^{+}$ ions in a rotating crystal. The beam position can be rapidly steered along a linear path between the crystal centre and its boundary. Combined with the rigid-body rotation of the crystal at frequency $\omega_r/2\pi$, this provides access to individual ions throughout the crystal. Inset: Example of using this method to prepare the letters \textit{SYD} in the crystal. Image reconstructed in the rotating frame of the ion crystal.
        (b) The addressing-beam power $P$ is synchronously modulated so that $\delta_\mathrm{ACSS}$ is applied only during the transit of a selected ion through the beam, yielding a controlled local $R_z$ spin rotation.
        (c) Simplified level scheme of ${}^{9}\mathrm{Be}^{+}$ at $B=\SI{2}{\tesla}$. The qubit is encoded in the Zeeman sublevels $\ket{\downarrow} \equiv \ket{{}^{2}S_{1/2},m_j=-1/2}$ and $\ket{\uparrow} \equiv \ket{{}^{2}S_{1/2},m_j=+1/2}$, separated by approximately \SI{55}{\giga\hertz} and driven globally by microwaves (grey). Doppler cooling and state-dependent fluorescence detection (black) are performed with light near \SI{313}{\nano\metre}, for which $\ket{\uparrow}$ is resonant and scatters photons and $\ket{\downarrow}$ is dark. The linear polarised addressing beam (pink) is detuned by $\Delta=\SI{2}{\giga\hertz}$ below the $\ket{\downarrow}\rightarrow\ket{{}^{2}P_{3/2},m_j=+1/2}$ transition.
        (d) Minimal pulse sequence for preparing arbitrary product states with independently programmable Bloch-sphere angles $\theta_i$ and $\phi_i$ for each ion $i$. Starting from $\ket{\uparrow}^{\otimes N}$, a global $\pi/2$ pulse about $+x$ rotates all spins to the equatorial plane. Site-selective $R_z(\theta_i)$ operations (pink bars) imprint ion-dependent phases, which are rotated into polar angles by the subsequent global $\pi/2$ pulse about $-x$. Site-selective $R_z(\phi_i)$ operations then set the azimuthal angles, enabling arbitrary local single-qubit state preparation. Bloch spheres illustrate the spin evolution during each pulse segment, with initial states shown in blue, final states in green, and applied rotations in red.
    }
    \label{fig:SQR-overview}
\end{figure}

A schematic of the experimental setup and protocol for single-ion addressing is shown in Fig.~\ref{fig:SQR-overview}.
A tightly focused off-resonant laser beam of waist radius $w_0\approx\SI{13.5}{\micro\meter}$ is incident perpendicular to the crystal plane.
To address an individual ion, the beam position can be continuously steered along a linear path by varying the drive frequency of an acousto-optic modulator (AOM), see Fig.~\ref{fig:SQR-overview}a.
A second AOM is used for laser power modulation to generate a fixed $\tau=\SI{200}{\nano\second}$ duration addressing pulse, which is synchronised to the crystal rotation period, such that the beam overlaps with the target ion for the duration of the pulse, see Fig.~\ref{fig:SQR-overview}b. This protocol is akin to the standard radial-plus-timing control mechanisms used to access arbitrary data on a hard disk drive. A simplified level scheme and the involved transitions in ${}^{9}\mathrm{Be}^{+}$ are shown in Fig.~\ref{fig:SQR-overview}c.
Addressing multiple ions in the crystal is performed sequentially, therefore the addressing time scales linearly with the number of addressed ions. This differs from approaches based on globally applied spatially structured beams, where the addressing time is primarily limited by the available laser power and the crystal rotation period \cite{Jee2026,Polloreno2022}.

Differences in ion tangential velocity result in different intensity modulation due to the ion's motion during the addressing pulse. This is compensated by calibrating the optical power $P_\pi$ required for an $R_z(\pi)$ gate as a function of radius. Arbitrary rotations $R_z(\theta)$ are implemented by scaling the power to $(\theta/\pi)P_\pi$ while keeping the pulse duration fixed. Larger rotations are realised using multiple pulses over successive crystal rotations.

The programmable local spin rotations $R_z(\theta_i)$ created by the addressing beam can be combined with global spin rotations to apply arbitrary local spin rotations to each ion spin $i$ in the crystal.
This is shown in Fig.~\ref{fig:SQR-overview}d, where two pulse sequences for local spin rotations are interleaved with global spin rotations, giving independent control of the two Bloch-sphere angles, enabling the preparation of arbitrary product states.

Compared with the previously used radius-selective optical pumping protocol \cite{Jee2026}, the present implementation accounts for the position of every individual ion. To achieve this, a reference ion crystal image, as shown in Fig.~\ref{fig:SQR-performance}b, is taken at the beginning of each experimental iteration during Doppler cooling when all ions are bright. This image is processed to determine the instantaneous ion coordinates, which are used to generate the two arbitrary waveforms applied to the steering and power-control AOMs. The full update takes approximately \SI{60}{\milli\second}, comparable to the total detection time. Repeating this shot-by-shot waveform generation reduces addressing errors arising from crystal rearrangements. Further details of the optical setup, beam alignment, and calibration procedures are provided in the Supplemental Material.

\section{Performance}
\label{sec:performance}

\begin{figure}[h!tb]
    \centering
    \includegraphics{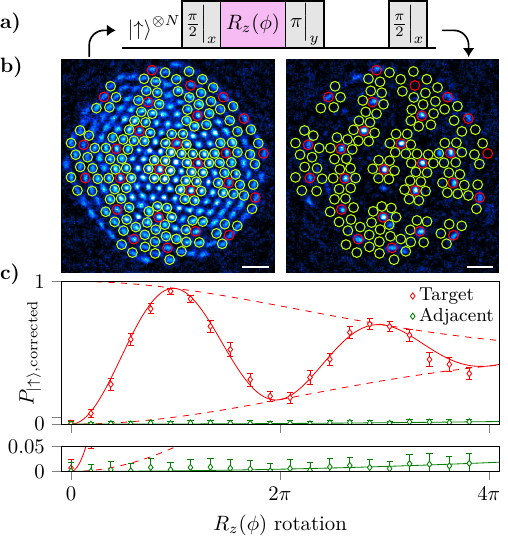}
    \caption{
        Evaluation of the addressing performance.
        (a) Ramsey spin-echo sequence used to characterise targeted spin rotations and crosstalk to neighbouring ions. During Doppler cooling and initialisation in $\ket{\uparrow}^{\otimes N}$, a reference image is acquired. A global $\pi/2$ pulse about the $x$ axis prepares all spins in an equal superposition state, after which individually targeted $R_z(\phi)$ phase rotations are applied. A global $\pi$ pulse together with a free-precession interval forms a spin echo that compensates for slow magnetic-field drift-induced phase rotations. A final global $\pi/2$ pulse about the $x$ axis maps the accumulated phase onto spin population for fluorescence detection.
        (b) Reference image (left) showing spatially well-separated target ions (red) and neighbouring ions (green), together with an image (right) obtained after programming a target rotation of $R_z(\pi)$. Scale bars: \SI{50}{\micro\metre}.
        (c)  SPAM-corrected bright-state probability $P_{\ket{\uparrow},\mathrm{corrected}}$ for target ions (red) and neighbouring ions (green) as a function of the programmed phase rotation $R_z(\phi)$. Error bars denote the standard error of the mean from 50 repetitions. The red solid line is a fit to the target-ion Ramsey fringe, while the red dashed lines indicate the fitted Gaussian dephasing envelope. The green solid line shows the corresponding fit for neighbouring ions.
    }
    \label{fig:SQR-performance}
\end{figure}

We quantify the addressing performance by measuring both the fidelity of local single-qubit $R_z$ phase gates and the residual crosstalk to neighbouring ions, as shown in Fig.~\ref{fig:SQR-performance}. The measurement sequence, illustrated in Fig.~\ref{fig:SQR-performance}a, begins by preparing all ions in $\ket{\uparrow}$ \cite{Wolf2024}. During this preparation, a reference image is acquired to identify a set of spatially well-separated target ions together with their nearest neighbours, Fig.~\ref{fig:SQR-performance}b. A Ramsey spin-echo sequence is then used to characterise the site-selective $R_z(\phi)$ gates applied only to the target ions during the first arm of the sequence. The final $\pi/2$ pulse maps the accumulated phase of both target and neighbouring ions onto their bright-state probabilities.

Scanning the programmed phase $\phi$ by varying the addressing laser power yields Ramsey fringes for the target ions and much slower residual oscillations on the neighbouring ions, see Fig.~\ref{fig:SQR-performance}c. After correcting for the independently measured SPAM infidelity, the contrast loss of the target-ion fringe is fitted with a Gaussian envelope, $W(\phi)\propto \exp\!\left[-\phi^2/(2\sigma^2)\right]$, with $\sigma=2.2(1)\pi$ equating to a fidelity of $0.950(4)$ for an $R_z(\pi)$ rotation. The inhomogeneous dephasing is the result of imperfect phase accumulation due to shot-to-shot variations in the local AC Stark shift, arising from finite in-plane ion motion, residual beam-pointing fluctuations, and fluctuations in addressing-laser power. Off-resonant scattering from the addressing beam leads to an exponential loss in fidelity, which is estimated to be $0.004$ at the present detuning of \SI{2}{\giga\hertz}.

Crosstalk arises from a nonzero overlap between adjacent ions and the addressing laser.
The ratio of the fitted neighbouring-ion to target-ion oscillation rate yields a crosstalk error of $0.021(6)$, which is determined primarily by two geometric factors.
First, the ratio of average ion-ion distance to the Gaussian laser beam waist, $s/w_0 \approx 1.6$, results in a finite residual laser intensity for neighbouring ions.
Second, the motion of the beam in the co-rotating frame causes an effective broadening of the beam profile (Fig.~\ref{fig:SQR-overview}b), which increases with ion radius, leading to crosstalk with adjacent ions at a similar radius as the target ion.
For the present crystal parameters, a pulse duration of $\tau = \SI{200}{\nano\second}$ provides a balance that avoids increased crosstalk at large radii while remaining long enough to keep the required optical power moderate.
More detailed modelling of both crosstalk and infidelity, along with mitigation strategies, is presented in the Supplemental Material.

\section{Applications of Coherent Site-Selective Spin Control}
\label{sec:demonstration}

Programmable single-qubit rotations enable the deterministic preparation of spatially structured product states with applications in quantum simulation and sensing. Here we demonstrate this capability in three settings: First, layer-selective addressing in a partially bilayer crystal, applicable to multilayer simulation and sensing \cite{Hawaldar2024}, second, preparation of a nontrivial biskyrmion spin texture with winding number of 2, potentially relevant to topological spin-texture dynamics and devices \cite{Kaneda2016,Shengbin2023,Psaroudaki2021,Koshibae2015,Luo2021}, and third, dual-quadrature Ramsey sensing in which different regions of the crystal simultaneously probe orthogonal quadratures of an accumulated Ramsey phase, extending the dynamic range from $\pi$ to $2\pi$ \cite{Zheng2024}.

\begin{figure}[t!tb]
    \centering
    \includegraphics{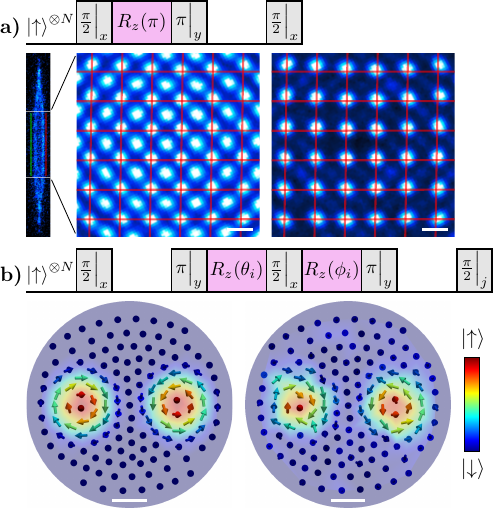}
    \caption{
        Experimental demonstrations of programmable site-resolved single-spin control in rotating ion crystals.
        (a) Layer-selective addressing in a partially bilayer crystal. Ramsey spin-echo sequence used to address all ions in one layer of a bilayer crystal using site-selective $R_z(\pi)$ gates. The left image shows a radial image of the crystal. The middle shows a reference image of the bilayer region consisting of two axially offset square lattices with the fitted lattice overlaid (red). The right shows the corresponding state-preparation result after applying the addressing sequence. Images are averaged over 50 repetitions. Scale bars: \SI{20}{\micro\metre}.
        (b) Preparation of a nontrivial spin texture. Ramsey spin-echo pulse sequence used to combine two rounds of site-selective $R_z$ gates with global microwave rotations, enabling independent control of the two Bloch-sphere angles for each ion.
        Target (left) and experimentally reconstructed (right) biskyrmion spin texture in a crystal containing approximately 150 ions. Arrows indicate the local Bloch vector, while the colour scale shows $\langle \sigma_z\rangle$. Scale bars: \SI{50}{\micro\metre}.
    }
    \label{fig:demons-sim}
\end{figure}

First, we apply site-selective $R_z(\pi)$ rotations to a partially bilayered crystal, as shown in Fig.~\ref{fig:demons-sim}a. In the bilayer region, the ions form two offset square lattices, with each layer occupying the plaquettes of the other \cite{Mitchell1998}. The projected ion positions are non-overlapping, allowing the two layers to be resolved and addressed independently. Moreover, the axial layer separation of $\sim\SI{50}{\micro\metre}$ is small compared with the addressing-beam Rayleigh range of $\sim\SI{2}{\milli\metre}$, therefore both layers effectively experience the same beam intensity.
The state-preparation procedure includes fitting two offset square lattices to the ion crystal.
Applying $\pi$ phase shifts to all ions in one layer and then rotating all ions into the $\sigma_z$ basis prepares one layer to be bright and the other to be dark, thereby demonstrating layer-selective addressing.
For the phase shifted layer a SPAM-corrected fidelity of $0.936(9)$ was measured, comparable to the $R_z(\pi)$ gate fidelity measured for a single-layer crystal in Fig.~\ref{fig:SQR-performance}.
For the phase unshifted layer a fidelity of $0.986(4)$ was measured, from which we calculate a crosstalk of $0.019(2)$.

Next, we prepare a biskyrmion spin texture using a Ramsey spin-echo sequence with two rounds of site-selective $R_z$ gates, combined with global microwave pulses (see Fig.~\ref{fig:demons-sim}b). As described in Ref.~\cite{Jee2026}, the local spin state is reconstructed from SPAM-corrected measurements in three global analysis bases, illustrated by the $\pi/2$ analysis pulse around axis $j=\{x,y,z\}$ in Fig.~\ref{fig:demons-sim}b. This allows the Bloch vector $\vec{s}_i=(\langle \sigma_x\rangle_i,\langle \sigma_y\rangle_i,\langle \sigma_z\rangle_i)$ to be determined for the $i$-th ion. The local fidelity is then obtained from the overlap of the reconstructed and target Bloch vectors, $\mathcal{F}_i=(1+\vec{s}_i\cdot \vec{s}_{i,\mathrm{tar}})/{2}$.
Using this procedure, we obtain a mean local fidelity of $\overline{\mathcal{F}}_i=0.92(2)$ for the biskyrmion.
The fidelities for these multistep state-preparation sequences are lower than the single-gate benchmark of Fig.~\ref{fig:SQR-performance}, primarily because the two sequential addressing operations, $\theta_i\in[0, \pi]$ and $\phi_i\in[0,2\pi]$, require a larger total rotation angle, increasing dephasing error.
The achieved fidelity is comparable to that reported in Ref. \cite{Jee2026} for a skyrmion created using globally applied spin-dependent forces that generate a spatially varying spin-motion coupling.

\begin{figure}[h!tb]
    \centering
    \includegraphics{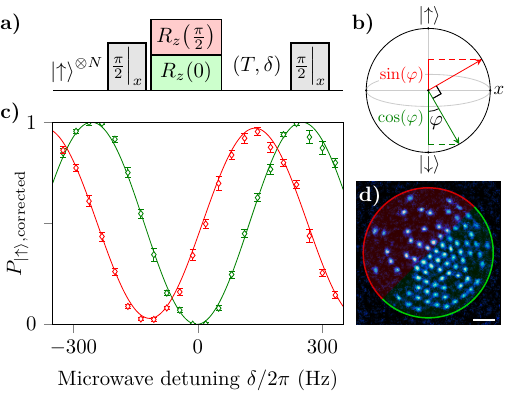}
    \caption{
        Dual-quadrature sensing demonstration.
        (a) Ramsey sequence for spatially resolved quadrature readout. Following initialisation in $\ket{\uparrow}^{\otimes N}$ and a global $\pi/2$ pulse, the addressing beam applies a local $R_z(\pi/2)$ rotation to one half of the crystal. During the free-evolution interval $T$, a phase $\varphi=\delta T$ accumulates for a microwave detuning $\delta$. A final global $\pi/2$ pulse maps the phase onto spin populations.
        (b) Bloch-sphere representation of the quadrature encoding. The two spatial regions probe orthogonal quadratures of the accumulated phase, yielding signals proportional to $\cos\varphi$ and $\sin\varphi$.
        (c) Ramsey spectroscopy for $T\approx\SI{2}{\milli\second}$. Green and red points show the SPAM-corrected bright-state population measured in the two halves of the crystal, corresponding to the cosine and sine quadratures, respectively. Solid lines are sinusoidal fits. Error bars denote the standard error of the mean.
        (d) Fluorescence image illustrating the two spatial sensing regions with different average bright-state populations ($\delta/2\pi=\SI{300}{Hz}$). Scale bar: $\SI{50}{\micro\meter}$.
    }
    \label{fig:demos-sensing}
\end{figure}

In addition to deterministic state preparation for quantum simulation, local control enables simultaneous measurement of orthogonal Ramsey quadratures. We demonstrate this capability through dual-quadrature Ramsey interferometry.

In a conventional Ramsey sequence consisting of a $\pi/2$ pulse, a free-precession interval $T$, and a second $\pi/2$ pulse, the measured signal is proportional to $\cos\varphi$, where $\varphi=\delta T$ is the phase accumulated for a frequency detuning $\delta$.
Because the cosine signal is symmetric about $\varphi=0$, the measured population alone does not uniquely determine the sign of the accumulated phase, leading to ambiguities modulo $\pi$ and limiting the unambiguous sensing range to $\varphi\in[0,\pi]$. Here, by simultaneously measuring signals proportional to both $\cos\varphi$ and $\sin\varphi$, the phase ambiguity is removed and the unambiguous sensing range is extended to $\varphi\in[-\pi,\pi)$.

The sensing protocol is illustrated in Fig.~\ref{fig:demos-sensing}a. Following a global $\pi/2$ pulse, the addressing beam applies a local $R_z(\pi/2)$ rotation to one half of the crystal, thereby introducing a relative phase offset of $\pi/2$ between the two spatial regions. During a free precession interval $T\approx\SI{2}{\milli\second}$, the spins accumulate the Ramsey phase $\varphi=\delta T$ due to an intentional microwave detuning $\delta$. A final global $\pi/2$ pulse maps the accumulated phase onto spin population, yielding simultaneous measurements proportional to $\cos\varphi$ and $\sin\varphi$ in the two spatial regions of the crystal, see Fig.~\ref{fig:demos-sensing}b.

From the sinusoidal fits to the Ramsey fringes in Fig.~\ref{fig:demos-sensing}c, we extract contrasts of $C=1.001(15)$ and $C=0.942(16)$ for the regions measuring $\cos(\varphi)$ (green) and $\sin(\varphi)$ (red), respectively, see Fig.~\ref{fig:demos-sensing}d. The reduced contrast in the $\sin(\varphi)$ region arises from the additional single-ion addressing operation applied to that part of the crystal and the reduction is consistent with a local $R_z(\pi/2)$ gate fidelity of $0.971(8)$.

\section{Conclusion and Outlook}

In this work, we demonstrated programmable coherent single-ion addressing in rotating Penning-trap crystals. Using site-selective AC Stark shifts, we prepared spatially structured spin states, performed layer-selective operations in bilayer crystals, and demonstrated dual-quadrature Ramsey sensing.
These results extend quantum control in large multidimensional ion crystals beyond globally driven spin rotations.

The capability to prepare spatially structured spin states, in combination with spin-dependent optical dipole forces or M{\o}lmer--S{\o}rensen interactions, enables studies of nonequilibrium dynamics seeded by engineered domain walls, vortices, and topological spin textures. This is directly relevant to proposed Penning-trap simulations of chiral $p+ip$ superconductors \cite{Shankar2022a} and holds potential for applications in spintronics for exploring skyrmion collision dynamics \cite{Kaneda2016, Shengbin2023}, skyrmion qubits \cite{Psaroudaki2021}, and magnetic memory devices \cite{Koshibae2015, Luo2021}.
The same techniques naturally extend to multilayer crystals, where site and layer-selective rotations and readout provide new quantum sensing and simulation capabilities. Combined with tunable interlayer interactions, this could enable measurements of interlayer correlations, bipartite entanglement, and transport between spatially separated ion layers \cite{Hawaldar2024}.
In addition, the ability to simultaneously sense different spin components by dividing the ion crystal into different regions may enhance the development of quantum sensing protocols to search for beyond-standard-model physics \cite{Zheng2024, SinananSingh2024, Budker2022, Carney2021a}.

Future improvements in fidelity could be achieved through enhanced planar-mode cooling \cite{Johnson2024,Johnson2025}, reduced addressing-beam waist and pointing fluctuations, and composite or optimal-control pulse sequences. Extending the approach to spatially selective entangling interactions would further expand Penning-trap crystals from globally interacting spin ensembles toward programmable multidimensional quantum simulators.

\section*{Acknowledgements}
We would like to thank Christophe Valahu and Ting Rei Tan for valuable feedback on the manuscript.
This material is based upon work supported by the Air Force Office of Scientific Research (FA2386-23-1-4067), the Australian Research Council Centre of Excellence for Engineered Quantum Systems (CE170100009) and a private grant from H. and A. Harley.
We acknowledge support from the Sydney Quantum Academy (J.Y.Z.J., J.H.P., and G.C.) and the Australian Government Research Training Program (RTP) Scholarship (J.Y.Z.J., J.H.P., N.M.). A.S. acknowledges support by the Department of Science and Technology, Govt. of India through the INSPIRE Faculty Award (DST/INSPIRE/04/2023/001486), by the Anusandhan National Research Foundation (ANRF), Govt. of India through the Prime Minister’s Early Career Research Grant (PMECRG) (ANRF/ECRG/2024/001160/PMS) and by IIT Madras through the New Faculty Initiation Grant (NFIG).

\bibliography{references}

\section{Supplemental material}

\section{Experimental Setup}
\label{sup:exp-setup-extend}
\label{sec:optical_setup}

The addressing implementation for single-qubit phase gates relies on three technical advances compared to our previous setup for radius-selective optical pumping, described in Ref. \cite{Jee2026}: (i) correction of beam astigmatism to reduce the focused waist, (ii) crystal rotation synchronised beam steering and power modulation enabled by an arbitrary waveform generator (AWG)
(iii) a real-time software pipeline that localises ions, assigns target phases, compiles RF waveforms, and uploads them before the next experimental iteration. In the following, we will explain the experimental addressing setup and sequence in detail.

\subsection{Addressing optical system}

\begin{figure*}[h!tb]
    \centering
    \includegraphics[width=0.95\textwidth]{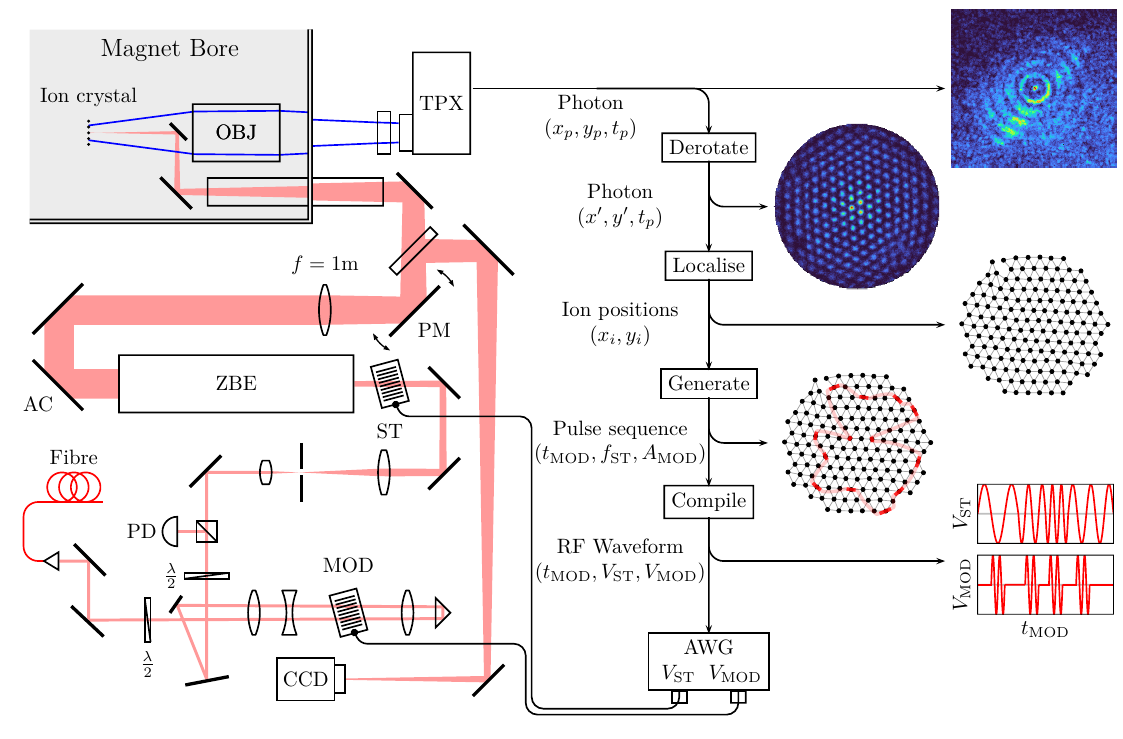}
    \caption{
        Illustration of the optical addressing setup and data processing.
        The left part of the figure shows a schematic of the addressing-beam delivery path. The fibre-delivered \SI{313}{nm} beam is power-modulated (MOD) via an AOM, spatially filtered by a pinhole, radially steered (ST) via an AOM, expanded by a zoom beam expanded (ZBE), corrected for residual astigmatism (AC), and focused towards the ion crystal by a \SI{1}{\metre} focal length lens. A piezo-actuated mirror (PM) enables precise beam alignment, while a weak pick-off beam is sent to a CCD camera for mode and position monitoring. The transmitted beam enters the magnet bore and is reflected by in-vacuum optics onto the ion crystal. Simplified photon detection: Ion fluorescence photons are collected by an imaging objective (OBJ) and detected by a Timepix3 camera (TPX). See the text for more details.
        The processing pipeline used to generate the AOM addressing signals is shown on the right. Photon positions and arrival times, $(x_p,y_p,t_p)$, recorded by the TPX camera are transformed (derotated, resulting in $(x^\prime,y^\prime,t_p)$) into the co-rotating frame and binned to form a reference image. Ion positions $(x_i,y_i)$ are extracted from this image and used to generate the scheduled addressing pulses, specified by their times $t_\mathrm{MOD}$, steering frequencies $f_\mathrm{ST}$, and modulation amplitudes $A_\mathrm{MOD}$. These pulse parameters are compiled into AWG voltage waveforms, $V_\mathrm{ST}$ and $V_\mathrm{MOD}$ for the steering and modulation AOMs, respectively.
    }
    \label{fig:optical-setup}
\end{figure*}

The addressing optical system, shown schematically in the left part of Fig.~\ref{fig:optical-setup}, is designed to deliver a diffraction-limited, steerable, pulsed beam with a waist smaller than the ion--ion spacing. The Penning trap is located approximately \SI{780}{\milli\metre} along the optical path inside the bore of a superconducting magnet \cite{Ball2019}, where no focusing optics can be placed. The addressing optics must therefore generate a tightly focused beam at the ion crystal using only components outside the magnet bore.

The optical path consists of four functional sections: fast power modulation, spatial filtering, radial beam steering and focusing, and diagnostic monitoring. Continuous-wave \SI{313}{nm} light is delivered through a polarisation-maintaining fibre \cite{Marciniak2017}. The beam first passes through a double-pass modulation AOM, which controls the optical power in the first diffracted order. The AOM is configured for optical rise and fall times of approximately \SI{40}{ns} for a focused beam diameter of about \SI{200}{\micro\metre}, enabling the $\tau=\SI{200}{ns}$ addressing pulses used in this work.

The beam is then passed through a $\times 3$ expanding spatial-filtering telescope. A pinhole at the intermediate focus suppresses higher-order spatial components, improving the Gaussian mode profile and reducing intensity tails that would otherwise contribute to crosstalk. The collimated beam exiting the telescope has a waist of approximately \SI{1}{\milli\metre}.

Radial steering is provided by an AOM with an active optical aperture of approximately \SI{2}{mm}, chosen to accommodate the expanded beam. The acoustic transit time across this aperture gives a frequency-switching time of approximately \SI{400}{ns}, which sets the dominant reconfiguration time for switching between ions at different radii. After the steering AOM, a diffraction-limited zoom beam expander increases the beam waist by a factor of approximately $12$, giving a \SI{12}{mm} waist before the focusing lens and a focused-beam steering range of approximately \SI{250}{\micro\metre} at the ion plane.

The expanded beam is focused with an $f=\SI{1}{m}$ lens. For the measured input beam size, Gaussian-beam propagation gives a diffraction-limited waist on the order of \SI{12}{\micro\metre}, consistent with the measured addressing waist. A piezo-actuated mirror placed after the focusing lens provides fine transverse position control for beam alignment at the ion crystal. A weak pick-off before the magnet bore directs a part of the beam to a CCD camera at a propagation distance equivalent to the ion crystal, allowing the beam position and mode profile to be monitored without interrupting the experiment.

The large beam diameter required for diffraction-limited focusing made the system sensitive to weak stress-induced surface deformations of the optical elements. Mechanical mounting of the \SI{2}{in} steering optics produced measurable astigmatism, visible as different focal positions along the two transverse beam axes and an increased effective waist at the ion plane. We compensated this residual astigmatism with a \SI{2}{in}-diameter, \SI{5}{mm}-thick flat mirror used as an adjustable weak cylindrical optic. By varying the mirror clamping pressure, we introduced a tunable curvature along one transverse axis; mounting the mirror on a rotation stage allowed this axis to be aligned with the measured astigmatism. This correction reduced the measured addressing waist from approximately \SI{17}{\micro\metre} to approximately \SI{13.5}{\micro\metre}.

\subsection{Ion localisation and addressing waveform generation}

Site-selective addressing requires knowledge of the instantaneous ion positions in the rotating-frame crystal. Although the equilibrium crystal configuration is set by the trapping potentials, the rotation frequency, and the rotating-wall potential, the lattice can change during an experiment. Background-gas collisions can convert $\mathrm{Be}^+$ into molecular ions such as $\mathrm{BeH}^+$ and $\mathrm{BeOH}^+$, which segregate towards the crystal boundary and can induce rearrangements of the remaining $\mathrm{Be}^+$ ions \cite{Sawyer2015}. Additional rearrangements can occur due to residual lattice stress generated by the radial Doppler-cooling laser and the rotating wall drive. To minimise addressing errors due to such changes, we acquire a reference image at each experimental iteration to inform updates to the addressing waveforms.

The processing sequence is illustrated in Fig.~\ref{fig:optical-setup}. During the initial Doppler-cooling interval, fluorescence from the crystal is collected by the objective lens and detected with a Timepix3 camera. The camera records the position and arrival time of each photon and streams them to a computer for processing. Using the measured crystal centre and the known crystal rotation frequency, each photon coordinate is transformed from the laboratory frame into the co-rotating frame. This derotation can be applied directly to the photon data stream, resulting in negligible delay. Binning the transformed photon coordinates produces a rotating-frame image in which individual ions are spatially resolved \cite{Wolf2024}.

From this co-rotating frame image, the ion positions are extracted using a deterministic image-processing algorithm. Initial position estimates are obtained from local maxima. Each estimate is then refined by iteratively computing the first moment of the photon counts within a small region around the current position and updating the estimate towards this centroid. The iteration terminates when the local photon distribution is centred on the ion position. Estimates that converge to the same location are subsequently merged, yielding a list of ion coordinates. The localisation step takes approximately \SI{5}{ms} for a crystal of 150 ions, with the runtime scaling approximately linearly with the number of ions.

The measured ion coordinates are then converted into a target-ion addressing-pulse sequence. Each addressed ion is illuminated by a \SI{200}{ns} pulse generated with the modulation AOM, whose optical rise time is approximately \SI{40}{ns}. Before each pulse, the steering AOM sets the beam radius; its \SI{400}{ns} switching time sets the dominant reconfiguration time between ions at different radii. Including the optical pulse duration and timing margins, the minimum separation between the start times of consecutive addressing pulses is approximately \SI{640}{ns}.

To schedule the addressing pulses, the azimuthal angle of each target ion is calculated in the rotating frame. The minimum temporal separation between pulses is converted into an angular separation using the crystal rotation frequency. A greedy scheduling algorithm then orders the target ions to minimise the total rotation angle, and hence the total addressing duration, subject to the above-mentioned timing constraints. For the crystals used here, the total addressing time remains below $N\times\SI{1}{\micro s}$, where $N$ is the number of addressed ions. If the available optical power is insufficient to realise the desired phase rotation in a single transit, the rotation is split across multiple addressing pulses, increasing the total addressing time linearly. Pulse-sequence generation takes \SIrange{1}{2}{ms}, depending on the number of pulses.

For each scheduled pulse, the target-ion radius determines the steering-AOM frequency, while the programmed phase rotation determines the modulation-AOM amplitude. The resulting list of pulse times, steering frequencies, and modulation amplitudes is compiled into voltage-sample arrays for the two AWG channels controlling the steering and modulation AOMs. The AWG output is triggered by a logic signal synchronised to the crystal rotation, producing the RF waveforms that steer the addressing beam and modulate its power at the required times. Waveform compilation is performed on a GPU and is limited primarily by data transfer to the AWG. Uploading the samples for a \SI{1}{ms} addressing waveform takes approximately \SI{20}{ms}.

The full update, including ion localisation, pulse scheduling, waveform compilation, and AWG upload, can be completed in approximately \SI{60}{ms}. This can be improved further by pipelining the entire update process. To achieve this, a reference image acquired in iteration $n$ is used to generate the addressing waveform for iteration $n+1$. The waveform for the current iteration is therefore available immediately after Doppler cooling, while the processing for the next iteration occurs in parallel with the remainder of the experimental cycle. Since one experimental iteration takes approximately \SI{120}{ms}, such a pipelined implementation would add no additional overhead to the sequence.

\subsection{Alignment Procedure}
\label{sec:alignment}

\begin{figure}[tb]
    \centering
    \includegraphics{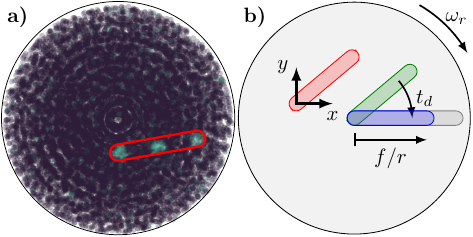}
    \caption{Alignment of the addressing beam.
        a) Example Ramsey sequence image used for coarse alignment. The addressing beam was programmed to apply approximately $R_z(\pi)$ at three positions along its trajectory, mapping the illuminated ions to $\ket{\uparrow}$ (light green) while unaddressed ions are rotated to $\ket{\downarrow}$ (black). The image is averaged over 75 shots. The red line indicates the addressing beam trajectory before alignment. The crystal diameter is approximately \SI{300}{\micro m}.
        b) Schematic of the three geometric calibration steps used to align the programmed addressing trajectory with the ion crystal. First, the piezo-actuated mirror translates the trajectory so that one end passes through the crystal centre (red $\rightarrow$ green). Second, the global timing offset $t_d$ is adjusted to rotate the trajectory in the crystal frame by $\omega_r t_d$ (green $\rightarrow$ blue). Third, the steering-AOM frequency-to-radial displacement calibration is adjusted (blue $\rightarrow$ grey).
    }
    \label{fig:SQR-alignment}
\end{figure}

Accurate alignment of the addressing beam trajectory is required to maximise the phase accumulated by target ion spins and minimise crosstalk to neighbouring ions. The alignment procedure, as illustrated in Fig.~\ref{fig:SQR-alignment}, calibrates the addressing beam trajectory, the timing of the addressing sequence relative to the crystal rotation, the conversion between steering-AOM frequency and addressing beam position displacement, and the addressing-beam-spot size.

Coarse alignment was performed with a Ramsey sequence as in Fig.~\ref{fig:demons-sim}a in which the addressing beam was set to drive an approximately $R_z(\pi)$ rotation.
By randomising the initial crystal phase over repeated shots, the ions sample different laboratory-frame positions, providing a beam-profile measurement with an effective spatial resolution below the ion--ion spacing in accumulated images in the co-rotating frame, see Fig.~\ref{fig:SQR-alignment}a.
The final analysis pulse maps the spins onto the $\sigma_z$ basis, thereby allowing the addressing beam profile to be imaged via ion fluorescence.
This measurement was used to calibrate the addressing beam, as described below.

The geometric calibration consists of three steps, illustrated in Fig.~\ref{fig:SQR-alignment}b. First, the piezo-actuated mirror (see Fig~\ref{fig:optical-setup}) is used to translate the addressing trajectory so that the endpoint aligns with the crystal centre. The addressing-trajectory length is given by the steering-AOM-frequency range. The endpoint is set by the lowest steering-AOM frequency used to deflect the beam.
Second, the global delay $t_d$ of the addressing sequence relative to a specific crystal rotation phase is adjusted. Because the crystal rotates at angular frequency $\omega_r$, this delay sets the azimuthal angle of the addressing trajectory in the rotating frame. Third, the steering-AOM frequency scale is calibrated by measuring beam displacement as a function of applied frequency, thereby providing the conversion between target-ion orbit radius and steering frequency.

Fine alignment was performed using the addressing-performance sequence described in Fig.~\ref{fig:SQR-performance}. The residual phase acquired by neighbouring ions was measured as a function of target-ion orbit radius and angular position of the neighbouring ions. The angular dependence of this crosstalk provides a sensitive diagnostic of residual beam-position and timing errors, which were iteratively corrected. In future implementations, the same procedure could be supplemented by maximising the target-ion phase accumulation, or equivalently, minimising the optical power required for a fixed $R_z(\pi)$ rotation, at each radius.

Long-term beam drift was small compared with the ion--ion spacing. Over a three-day period we observed a drift below \SI{10}{\micro m}. Independent monitoring on the diagnostic CCD path showed short-term beam-pointing stability of approximately \SI{1}{\micro m}.

\section{Addressing Infidelity and Crosstalk}

\subsection{Geometric Model}

\begin{figure}[tb]
    \centering
    \includegraphics[width=8.5cm]{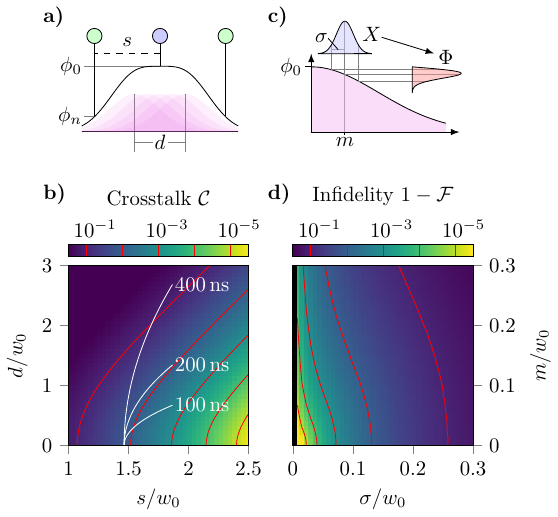}
    \caption{
        Geometric models for addressing crosstalk and target-ion infidelity.
        \textbf{a)} Crosstalk model. During a finite addressing pulse, the ion moves relative to the addressing beam in the co-rotating frame, so that the accumulated phase is determined by the time-integrated beam intensity. For a pulse in which the beam travels a distance $d$, this produces an effective broadened illumination profile. The crosstalk is defined as the relative accumulated phase $\phi_\mathrm{n}/\phi_0$ at a neighbouring ion separated from the target ion by a distance $s$.
        \textbf{b)} Calculated crosstalk $\phi_\mathrm{n}/\phi_0$ as a function of ion spacing $s$ and beam travel distance $d$, both normalised to the addressing beam waist $w_0$. Red contours indicate constant crosstalk. White curves, parametrised by the target ion radius, show the expected sampling of this parameter space for pulse durations of $\tau=100$, $200$, and $\SI{400}{\nano\second}$ using the crystal parameters of the main text.
        \textbf{c)} Target-ion infidelity model. Shot-to-shot fluctuations of the ion position relative to the addressing beam are modelled by a Gaussian distribution of width $\sigma$, with a static offset $m$ from the beam centre. Sampling the Gaussian beam profile over this position distribution produces a distribution of accumulated phases $\Phi$, leading to imperfect target rotations.
        \textbf{d)} Calculated target-ion infidelity $1-\mathcal{F}$ for an $R_z(\pi)$ rotation as a function of position uncertainty $\sigma$ and static misalignment $m$, both normalised to $w_0$. Red contours indicate constant infidelity.
    }
    \label{fig:crosstalk}
\end{figure}

The performance of site-selective phase control is characterised by two distinct error mechanisms: crosstalk to neighbouring ions and imperfect phase accumulation on the target ion. Crosstalk is primarily determined by the spatial overlap between the addressing beam and adjacent ions. Target-ion infidelity, by contrast, is dominated by shot-to-shot variations in the ion position relative to the addressing beam. In this section, we introduce simple geometric models for both effects and compare the expected parameter scalings to the main text.

For the crosstalk model (see Fig.~\ref{fig:crosstalk}a), we approximate the addressing beam by a Gaussian intensity profile with waist $w_0$. In the co-rotating frame, a pulse of duration $\tau$ causes the beam to move relative to the ions by a distance
\begin{equation}
    d(r)=v(r)\tau=\omega_r r\,\tau,
\end{equation}
where $r$ is the ion radius, $\omega_r$ is the crystal rotation frequency. We ignore effects from the finite rise time of the optical pulse. The calculations in Fig.~\ref{fig:crosstalk}b were performed for an adjacent ion located at the same radius as the target ion, aligning it with the motion of the addressing beam in the co-rotating frame.

The accumulated phase is proportional to the time-integrated intensity experienced during the pulse. Equivalently, the finite beam motion can be treated as a convolution of the Gaussian beam with a top-hat trajectory of length $d$, giving an effective broadened intensity profile. The neighbouring-ion crosstalk is then
\begin{equation}
    \mathcal{C} = \frac{\phi_\mathrm{n}}{\phi_0},
\end{equation}
where $\phi_\mathrm{n}$ is the phase accumulated by a nearest neighbour and $\phi_0$ is the phase accumulated by the target ion.

Across the crystal, the crosstalk depends on two competing radial trends. The ion-ion spacing $s(r)=s_0+s_r r^2$ increases with increasing radius $r$ \cite{Dubin2013}, thereby suppressing crosstalk. At the same time, the beam travel distance $d(r)$ also increases with radius, broadening the effective illumination profile and increasing crosstalk. We evaluate these radius-dependent effects using the measured crystal geometry and the beam waist $w_0=\SI{13.5}{\micro\meter}$, shown as white curves in Fig.~\ref{fig:crosstalk}b. For the parameters of Fig.~\ref{fig:SQR-performance}, a pulse duration of $\tau=\SI{200}{\nano\second}$ provides a useful compromise: it keeps the required optical power moderate while avoiding the increased crosstalk at higher ion radii expected for longer pulses at large radius, as can be seen for $\tau=\SI{400}{\nano\second}$. In contrast, a reduction to $\tau=\SI{100}{\nano\second}$ can significantly reduce the crosstalk at the boundary.
Operation at higher crystal rotation frequencies would increase the beam travel distance resulting in a shorter optimal pulse duration which would require increased laser power.

The same geometric nature of the crosstalk also makes it predictable. Since the crosstalk is determined by the measured ion positions and the calibrated addressing beam profile, it can, in principle, be incorporated into a crosstalk-aware pulse compilation. In this approach, the programmed addressing phases are adjusted so that the sum of the directly applied phase and the phases induced by neighbouring pulses yields the desired final phase pattern. Any common phase offset is equivalent to a global $R_z$ rotation and can be absorbed into the phase of a subsequent microwave pulse. For a triangular lattice with nearest-neighbour crosstalk $\mathcal{C}$, compensating the largest possible contribution from all six neighbouring ions increases the required target rotation by at most a factor of approximately $1+6\mathcal{C}$. For the measured crosstalk of $\mathcal{C}\simeq0.021$, this corresponds to a $\sim 1.1$ increase in the required rotation angle.

We next consider the target-ion infidelity. This error arises because the phase accumulated by the target ion depends on its instantaneous position relative to the addressing beam. The relevant position uncertainty includes thermal in-plane motion, residual beam-pointing fluctuations, and localisation errors in determining the ion position before the addressing pulse.

We model the shot-to-shot displacement of the target ion relative to the addressing beam as a two-dimensional Gaussian distribution (see Fig.~\ref{fig:crosstalk}c). The unknown, residual static beam misalignment $m$ is taken along one transverse direction, such that
\begin{equation}
    X\sim \mathcal{N}(m,\sigma^2),\qquad
    Y\sim \mathcal{N}(0,\sigma^2),
\end{equation}
and $\sigma$ is the shot-to-shot rms position uncertainty.

For each sampled position, the ion experiences a local intensity resulting in an accumulated phase
\begin{equation}
    \Phi(X,Y) = \phi_0 \exp\left[-\frac{X^2+Y^2}{2w_0^2}\right].
\end{equation}
The laser power is calibrated so that the mean accumulated phase matches the target rotation. For a nominal $R_z(\pi)$ pulse, the actual rotation angle is therefore
\begin{equation}
    \Theta = \pi \frac{\Phi(X,Y)}{\langle \Phi(X,Y)\rangle}.
\end{equation}
The corresponding mean target-state infidelity is
\begin{equation}
    1-\mathcal{F} = \frac{1+\langle\cos\Theta\rangle}{2}.
\end{equation}

The resulting infidelity map is shown in Fig.~\ref{fig:crosstalk}d. The model shows that the fidelity is highly sensitive to the ratio $\sigma/w_0$. It also shows that static misalignment increases the sensitivity to position fluctuations, because the ion samples a region of the Gaussian beam with a larger intensity gradient. For $w_0=\SI{13.5}{\micro\meter}$ and zero static misalignment,
increasing the effective position uncertainty from $\sigma=\SI{2.7}{\micro\meter}$ to $\SI{2.9}{\micro\meter}$ reduces the predicted $R_z(\pi)$ fidelity from $0.955$ to $0.944$, consistent with the range of Gaussian dephasing limited fidelities observed experimentally.

This simple model does not distinguish between the different physical sources of position uncertainty. Independent measurements of in-plane mode temperature, beam pointing stability, and localisation precision would be required to separate these contributions. Nevertheless, the scaling in Fig.~\ref{fig:crosstalk}d indicates clear routes for improving the addressing fidelity: reducing in-plane motion, improving beam pointing stability, improving ion localisation, and increasing the ratio $w_0/\sigma$ while maintaining sufficiently low crosstalk.
\par\vspace{\baselineskip}
Finally, the effect of target-ion infidelity can also be reduced at the pulse-sequence level by minimising the total addressed phase accumulated by each ion. For example, in the arbitrary product-state sequence of Fig.~\ref{fig:demons-sim}, the local rotations can be distributed across the two arms of the echo sequence. First, a global $\pi/2$ phase change is used to map the desired angle $\theta\in[0,\pi]$ to the range $\theta'\in[-\pi/2,\pi/2]$. Then, the new angle may be decomposed into positive and negative contributions, $\theta'=\theta_+ + \theta_-$, with $\theta_+\in[0,\pi/2]$ and $\theta_-\in[-\pi/2,0]$, which are applied in separate arms of the echo sequence.
This reduces the maximum phase acquired by any individual spin, therefore reducing the phase-dependent infidelity caused by position fluctuations.

\subsection{Off-resonant scattering rate calculations}

\begin{figure}[h!tb]
    \centering
    \includegraphics[width=8.5cm]{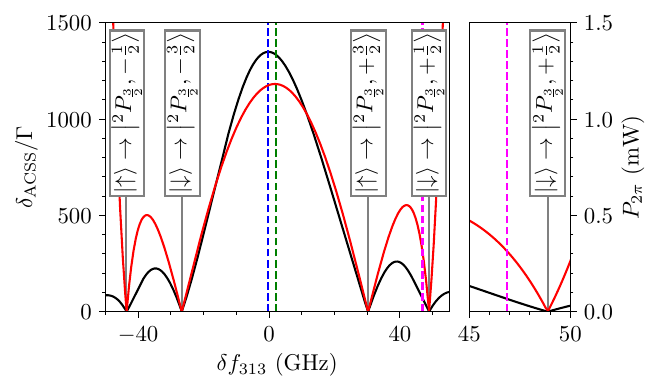}
    \caption{Addressing-laser detuning.
        Calculated differential AC Stark shift to off-resonant scattering ratio \(\delta_\mathrm{ACSS}/\Gamma\) (black, left axis) and optical power \(P_{2\pi}\) required for an \(R_z(2\pi)\) rotation in a \SI{200}{ns} pulse (red, right axis), plotted as a function of the addressing-laser detuning \(\delta f_{313}\) from the centre of the ${}^{2}S_{1/2}$ manifold to centre of the ${}^{2}P_{3/2}$ manifold laser frequency. The right panel zooms in on the detuning used in this work (magenta dashed line). The detuning giving the largest \(\delta_\mathrm{ACSS}/\Gamma\) ratio (blue dashed line) and the optical-dipole-force laser detuning (green dashed line) are shown for comparison.}
    \label{fig:frequency}
\end{figure}

The addressing beam implements local \(R_z\) rotations through the differential AC Stark shift \(\delta_\mathrm{ACSS}=\delta_{\mathrm{ACSS},\uparrow}-\delta_{\mathrm{ACSS},\downarrow}\) between the two qubit states. The same off-resonant coupling also produces photon scattering at a rate \(\Gamma\), which contributes an irreversible error to the addressing operation. The relevant figure of merit is therefore the ratio \(\delta_\mathrm{ACSS}/\Gamma\), which gives the coherent phase shift that can be accumulated per scattering event.
The exponential decay envelope from off-resonant scattering is given by
\begin{equation}
    W(\phi) = \exp\left(-\frac{\phi}{2\pi}\frac{\Gamma}{\delta_\mathrm{ACSS}}\right),
\end{equation}
from which the scattering contribution for an $R_z(\pi)$ rotation can be calculated as
\begin{equation}
    1 - \mathcal{F} = \frac{1-W(\pi)}{2}.
\end{equation}
In the experiments reported here, the addressing laser was detuned by \(\Delta=\SI{2}{GHz}\) below the \(\ket{\downarrow}\rightarrow\ket{{}^{2}P_{3/2},m_j=+1/2}\) transition, see Fig.~\ref{fig:SQR-overview}c. This detuning was chosen because it is compatible with the existing laser system, but it is not the detuning that minimises scattering for a fixed addressing phase. For an isolated transition, the AC Stark-shift-to-scattering ratio increases approximately linearly with detuning, while the optical power required to obtain a fixed Stark shift also increases. In the multilevel \({}^{9}\mathrm{Be}^{+}\) system, the relevant ratios are calculated by summing over the allowed couplings to the excited-state manifold.

Following Refs.~\cite{Carter2023,Uys2010}, we calculate both \(\delta_\mathrm{ACSS}/\Gamma\) and the power required to produce a fixed addressing phase as a function of the laser detuning \(\delta f_{313}\), defined relative to the centre of the \({}^{2}S_{1/2}\rightarrow{}^{2}P_{3/2}\) manifold. The calculation uses the experimental beam geometry with an addressing waist of \(w_0=\SI{13.5}{\micro m}\) and the beam incident normal to the crystal plane. The polarisation is linear, which is perpendicular to the quantisation axis, therefore capable of driving $\Delta m_j=\pm1$. The required power \(P_{2\pi}\) is the optical power at the ion plane needed to drive an \(R_z(2\pi)\) rotation in a \(\tau=\SI{200}{ns}\) pulse, corresponding to a differential Stark shift of \(\delta f_\mathrm{ACSS}=1/\tau=\SI{5}{MHz}\).

The results are shown in Fig.~\ref{fig:frequency}. At the detuning used in this work, \(\delta f_{313}\simeq\SI{47}{GHz}\), we find \(\delta_\mathrm{ACSS}/\Gamma\simeq65\), with a required power \(P_{2\pi}\simeq\SI{0.3}{mW}\). The largest ratio within the calculated range occurs near \(\delta f_{313}\simeq\SI{-0.2}{GHz}\), where \(\delta_\mathrm{ACSS}/\Gamma\simeq1350\) and \(P_{2\pi}\simeq\SI{1.2}{mW}\).
Moving to this detuning would reduce the scattering contribution to the addressing infidelity, $1- \mathcal{F}$, by approximately a factor of 20, from $0.004$ to $0.0002$, at the cost of a fourfold increase in optical power.

The detuning of the optical-dipole-force lasers, \(\delta f_{313}\simeq\SI{2}{GHz}\), gives a similar ratio, \(\delta_\mathrm{ACSS}/\Gamma\simeq1330\). This detuning is therefore also attractive for future addressing experiments, since it would allow the addressing beam to share laser light with this laser system while remaining within the available optical-power budget.
\vspace{-4pt}

\end{document}